\documentclass[twocolumn,showpacs,preprintnumbers,amsmath,amssymb,APS]{revtex4}

\usepackage{dcolumn}
\usepackage{bm}
\usepackage[spanish,english]{babel}
\usepackage{amsfonts}
\usepackage{amssymb}
\usepackage{graphicx}
\usepackage[latin1]{inputenc}
\newcommand{\R}{\mathcal{R}}
\newcommand{\E}{\mathcal{E}}

\begin{document}

\title{ {\bf The Hydrogen atom in Palatini theories of gravity}}
\author{Gonzalo J. Olmo}\email{golmo@perimeterinstitute.ca}
\affiliation{ { Perimeter Institute, 31 Caroline St. N, Waterloo, ON N2L 2Y5, Canada}}

\pacs{98.80.Es , 04.50.+h, 04.25.Nx}

\date{23rd February, 2008}

\begin{abstract}
We study the effects that the gravitational interaction of $f(R)$ theories of gravity in Palatini formalism has on the stationary states of the Hydrogen atom. We show that the role of gravity in this system is very important for lagrangians $f(R)$ with terms that grow at low curvatures, which have been proposed to explain the accelerated expansion rate of the universe. We find that new gravitationally induced terms in the atomic Hamiltonian generate a strong backreaction that is incompatible with the very existence of bound states. In fact, in the $1/R$ model, Hydrogen disintegrates in less than two hours. The universe that we observe is, therefore, incompatible with that kind of gravitational interaction. Lagrangians with high curvature corrections do not lead to such instabilities.
\end{abstract}

\pacs{98.80.Es , 04.50.+h, 04.25.Nx}

\maketitle

\section{Introduction}

The accelerated expansion rate of the universe \cite{Tonry-Knop} is one of the biggest puzzles that theoretical physics faces nowadays. Dark energy sources within the framework of General Relativity (GR) have been postulated as the missing element that could explain that phenomenon. On the other hand, modified theories of gravity have been proposed as an alternative to dark energy sources. Modified theories usually provide self-accelerated cosmic solutions on purely geometrical grounds, making unnecessary the introduction of unobserved exotic sources of matter-energy. The modification of the gravitational laws is, however, a very delicate issue in which intuition is not always a good guide. In fact, a modification originally thought to affect the dynamics at large scales could end up having non-trivial effects at shorter scales. It is thus necessary to study the dynamics of the different gravitational theories in different regimes and identify their positive and negative aspects aiming at learning how to construct theories that exhibit the desired properties. In this sense, theories of gravity in which the lagrangian is some function $f(R)$ of the scalar curvature $R$ manifest many interesting properties and have attracted much attention in the recent literature. \\ 

The equations of motion of $f(R)$ theories can be derived in two different ways depending on whether the connection is seen as independent of the metric (Palatini formalism) or as dependent of it (metric formalism). In the metric formalism, i.e., when the connection is the Levi-Cività connection of the metric, besides the metric one identifies an additional scalar degree of freedom, which turns the scalar curvature $R$ into a dynamical object. The interaction range of this scalar field depends on the form of the lagrangian and can change due to different reasons. When perturbation theory is applicable, the first order approximation shows that, for models of interest in the late-time cosmic acceleration, the interaction range changes driven by the cosmic expansion\cite{Olmo07b}. The scalar field can be short-ranged for some time (radiation and matter dominated eras) and then turn into a long-ranged field, causing late-time cosmic acceleration. This type of lagrangian is ruled out by solar system experiments. Some $f(R)$ lagrangians, however, cannot be treated perturbatively. Nonetheless, one can still define an effective mass or interaction range for the scalar, which now depends on the local matter density. The constraints on such lagrangians by local experiments have been discussed recently in the literature\cite{chameleon}.\\ 

In this paper we will consider the other formulation of $f(R)$ theories, namely, the Palatini formalism. Almost surprisingly, allowing the connection to be determined by the equations of motion does not introduce new dynamical degrees of freedom. In these theories, the metric turns out to be the only dynamical field, which satisfies second-order differential equations. As we will see, the effect of the lagrangian $f(R)$ is to change the way matter generates the metric by introducing on the right hand side of the field equations new matter terms that depend on the trace of the energy-momentum tensor of the sources. In vacuum, the field equations reduce (always and exactly) to those of GR with a cosmological constant.  For this reason, it has been thought for some time that Palatini $f(R)$ theories could pass the solar system observational tests \cite{Vol03,solsys} (see also \cite{Sot06} for a discussion of this point). It has also been shown that different choices of lagrangian $f(R)$ are able to accommodate several of the different cosmic eras of the standard cosmological model \cite{Fay07}. However, a more careful analysis of the gravitational dynamics in the presence of sources indicates that these theories might be in strong conflict with our understanding of the microscopic \footnote{Note that we are assuming that the proposed $f(R)$ models are as fundamental as GR and, therefore, should provide a consistent description of Nature in all experimentally accessible scales. To argue that the correcting terms of the lagrangian are only applicable in cosmic scales should come with a detailed explanation of why at different scales a different gravity lagrangian should be used. In the absence of such an explanation, we treat the $f(R)$ lagrangians as fundamental and check their predictions at atomic scales, where GR predicts a virtually flat space-time structure.} world \cite{Olmo07,Olmo05,Fla04} (the study of polytropic matter configurations also points in this direction \cite{Sot07}). This aspect, together with the cosmological viability of different models, is of primary importance due to the fact that almost all cosmological observations relay on the detection of electromagnetic radiation, which is intimately related to the quantum mechanical nature of atomic and molecular structure. In this work, we elaborate in this direction and study how the gravitational interaction in Palatini $f(R)$ theories affects the  non-relativistic limit of the (one-particle) Dirac equation. The analysis of various gravitationally-induced correcting terms in the resulting Schrodinger-Pauli equation will provide us with solid arguments against the existence in the gravity lagrangian of correcting terms relevant at low cosmic curvatures. As we will see, the very low matter density (or curvature) scales that characterize (infrared-corrected) modified lagrangians can be reached near the zeros of the wavefunctions. This causes a strong gravitational backreaction that makes unstable the stationary states of the Hydrogen atom. In particular, we find that the ground state disintegrates in a matter of hours. On the contrary, if the gravity lagrangian is modified by high curvature corrections, the {\it backreaction} effects are negligible and the atom remains stable. \\ 

The paper is organized as follows. In section II, we define the action of $f(R)$ theories in Palatini, derive the field equations, and discuss the metric generated by microscopic systems. We then introduce two illustrative models, namely, $f(R)=R+\frac{R^2}{R_P}$ and $f(R)=R-\frac{\mu^4}{R}$, which will help us compare the behavior of the metric when the GR action gets ultraviolet or infrared corrections, respectively. In section III we derive the non-relativistic limit of Dirac's equation starting from its curved space-time formulation. We then discuss the effects induced by the modified Schrodinger-Pauli equation in the stationary solutions of the Hydrogen atom. We conclude with a summary and discussion of the results obtained. In the Appendix we estimate the decay rate of the ground state of the atom. 

\section{The theory}

Let us begin by defining the action of Palatini theories
\begin{equation}\label{eq:Pal-Action}
S[{g},\Gamma ,\psi_m]=\frac{1}{2\kappa^2}\int d^4
x\sqrt{-{g}}f({R})+S_m[{g}_{\mu \nu},\psi_m]
\end{equation}
Here $f({R})$ is a function of ${R}\equiv{g}^{\mu \nu }R_{\mu \nu }(\Gamma )$, with $R_{\mu \nu }(\Gamma )$ given by
$R_{\mu\nu}(\Gamma )=-\partial_{\mu}
\Gamma^{\lambda}_{\lambda\nu}+\partial_{\lambda}
\Gamma^{\lambda}_{\mu\nu}+\Gamma^{\lambda}_{\mu\rho}\Gamma^{\rho}_{\nu\lambda}-\Gamma^{\lambda}_{\nu\rho}\Gamma^{\rho}_{\mu\lambda}$
where $\Gamma^\lambda _{\mu \nu }$ is the connection. The matter action $S_m$ depends on the matter fields $\psi_m$, the metric $g_{\mu\nu}$, which defines the line element $ds^2=g_{\mu\nu}dx^\mu dx^\nu$, and its first derivatives (Christoffel symbols). The matter action does not depend on the connection $\Gamma^\lambda _{\mu \nu }$, which is seen as an independent field appearing only in the gravitational action (this condition is not essential and can be relaxed at the cost of introducing a non-vanishing torsion). Varying (\ref{eq:Pal-Action}) with respect to the metric $g_{\mu\nu}$ we obtain
\begin{equation}\label{eq:met-var-P}
f'(R)R_{\mu\nu}(\Gamma)-\frac{1}{2}f(R)g_{\mu\nu}=\kappa ^2T_{\mu
\nu }
\end{equation}
where $f'(R)\equiv df/dR$. From this equation we see that the scalar $R$ can be solved as an algebraic function of the trace $T$. This follows from the trace of
(\ref{eq:met-var-P})
\begin{equation}\label{eq:trace-P}
f'(R)R-2f(R)=\kappa ^2T,
\end{equation}
The solution to this algebraic equation will be denoted by $R=\R(T)$. The variation of (\ref{eq:Pal-Action}) with respect to $\Gamma^\lambda _{\mu \nu }$ must vanish
independently of (\ref{eq:met-var-P}) and gives
\begin{equation}\label{eq:con-var}
\nabla_\rho  \left[\sqrt{-g}\left(\delta ^\rho _\lambda
f'g^{\mu \nu }-\frac{1}{2}\delta ^\mu _\lambda f'g^{\rho
\nu }-\frac{1}{2}\delta^\nu_\lambda f'g^{\mu
\rho}\right)\right]=0
\end{equation}
where $f'\equiv f'(\R[T])$ is also a function of the matter terms. This equation leads to 
\begin{equation}\label{eq:Gamma-1}
\Gamma^\lambda_{\mu \nu }=\frac{t^{\lambda \rho
}}{2}\left(\partial_\mu t_{\rho \nu }+\partial_\nu
t_{\rho \mu }-\partial_\rho t_{\mu \nu }\right)
\end{equation}
where  $t_{\mu \nu }\equiv \phi g_{\mu \nu }$, and $\phi\equiv \frac{f'(\R[T])}{f'(\R[0])}$ is dimensionless and normalized to unity outside of the sources ($T=0$). It is now useful to rewrite (\ref{eq:met-var-P}) adding and subtracting $\frac{f'}{2}\R(T)g_{\mu\nu}\equiv
\frac{f'}{2}t^{\alpha\beta}R_{\alpha\beta}(\Gamma)t_{\mu\nu}$ to get 
\begin{equation}\label{eq:G-tmn}
f'G_{\mu\nu}(t)=\kappa^2T_{\mu\nu}-\frac{[\R f'-f]}{2\phi}t_{\mu\nu}
\end{equation}
where $G_{\mu\nu}(t)$ is the Einstein tensor associated to $t_{\mu\nu}$. The equations of motion (\ref{eq:G-tmn}) for the auxiliary metric $t_{\mu\nu}$ are considerably simpler than those for $g_{\mu\nu}$, 
\begin{eqnarray}\label{eq:Gmn}
R_{\mu \nu }(g)-\frac{1}{2}g_{\mu \nu }R(g)&=&\frac{\kappa
^2}{f'}T_{\mu \nu }-\frac{\R f'-f}{2f'}g_{\mu \nu
}-\nonumber\\&-&\frac{3}{2(f')^2}\left[\partial_\mu f'\partial_\nu
f'-\frac{1}{2}g_{\mu \nu }(\partial f')^2\right]+\nonumber \\
&+& \frac{1}{f'}\left[\nabla_\mu \nabla_\nu f'-g_{\mu \nu }\Box
f'\right] \ ,
\end{eqnarray}
because of the difficulty of dealing with the matter derivatives $\partial f'\sim \partial T$ and $\partial^2 f'\sim (\partial T)^2\sim (\partial^2T)$. 
Solving for $t_{\mu\nu}$ using the system (\ref{eq:G-tmn}) and then going back to $g_{\mu\nu}$ via the conformal transformation $g_{\mu\nu}=\phi(T)^{-1}t_{\mu\nu}$ is a useful simplification that makes the task of finding solutions much easier. Note that this fortunate circumstance is due to the fact that the conformal transformation completely cancels out the disturbing derivatives of (\ref{eq:Gmn}). Moreover, the conformal relation between the two metrics puts forward the fact that the metric $g_{\mu\nu}$ receives two kinds of contributions: non-local contributions that result from integration over the sources, which produce the term $t_{\mu\nu}$, and local contributions due to $\phi(T)$, which depend on the local details of $T$ at each space-time point. This local contribution arises due to the independent character of the connection and, to our knowledge, does not appear in any other metric theory of gravity, where the connection is generally assumed to be metric compatible (Levi-Cività connection). \\

The similarity between the field equations of GR and (\ref{eq:G-tmn}) suggests that for weak sources and reasonable choices of lagrangian $f(R)$ (those that lead to negligible {\it cosmological term} $\frac{\R f'-f}{2f'}$), the right hand side of (\ref{eq:G-tmn}) is small and, like in GR, $t_{\mu\nu}$ can be expressed as $t_{\mu\nu}(x)\approx \eta_{\mu\nu}+h_{\mu\nu}(x)$, where $|h_{\mu\nu}(x)|\ll 1$ is given as an integral over the sources (for details of exact calculations see \cite{Olmo07}). For very weak sources such as atoms or elementary particles, we find that the self-contribution to $h_{\mu\nu}\to 0$. On the other hand, if our microscopic system is placed in an external gravitational field whose contribution to $t_{\mu\nu}$ is not negligible, we can always take a coordinate system in which the metric at the boundaries of a {\it box} containing the system (large as compared to the microscopic system but small as compared to the range of variation of the external metric $t_{\mu\nu}$) becomes $\sim \eta_{\mu\nu}$. In both situations, the metric $g_{\mu\nu}$  becomes simply
\begin{equation}\label{eq:phi-eta}
g_{\mu\nu}(x)\approx\phi(T)^{-1}\eta_{\mu\nu}
\end{equation}
where $\phi(T)\to 1$ in the boundaries of our {\it auxiliary box} but might depart from unity within the sources, depending of the particular lagrangian chosen. \\
If one trivializes the role of the local term $\phi(T)$, then one finds that relative motion between particles is not very much affected by the modified gravity lagrangian \cite{LMS08}. However, we find that the presence of $\phi(T)^{-1}$ in front of $\eta_{\mu\nu}$ is very important because the matter fields in (\ref{eq:Pal-Action}) are coupled to $g_{\mu\nu}$ and, unlike in all other known metric theories of gravity, $g_{\mu\nu}$ only becomes locally $\eta_{\mu\nu}$ in regions where $T=0$ exactly. Consequently, the $\phi(T)$ dependence of $g_{\mu\nu}$ induces new interactions and self-interactions between the matter fields \cite{Olmo07}, as will be explained here in detail. 
Note that the presence of the (scalar) term $\phi(T)$ is physical and not a problem of choosing the {\it wrong} coordinate system, as criticized in \cite{Kai07}. In fact, if one computes geometrical invariants such as \footnote{Here ${R_{\alpha\beta\gamma}}^\lambda$ is the Riemann tensor of the metric $g_{\mu\nu}$.} ${R_{\alpha\beta\gamma}}^{\lambda}{R^{\alpha\beta\gamma}}_{\lambda}$, various derivatives of $\phi(T)$ appear and cannot be eliminated by choosing different coordinate systems because ${R_{\alpha\beta\gamma}}^{\lambda}{R^{\alpha\beta\gamma}}_{\lambda}$ is a coordinate invariant.  Note also that the dependence of ${R_{\alpha\beta\gamma}}^{\lambda}{R^{\alpha\beta\gamma}}_{\lambda}$ on the local energy-momentum distribution via $\phi(T)$ tells us that the geometry might be subjected to microscopic fluctuations driven by the fluctuations of $T$ and {\it modulated} by the form of the gravity lagrangian [recall that $\phi(T)\equiv f'(\R[T])/f'(\R[0])]$. Therefore, lagrangians sensitive to low energy scales could lead to unnacceptable microscopic curvature fluctuations, while others could lead to more robust geometries which would only fluctuate at very high energies. In this latter case, however, neglecting the contribution of $h_{\mu\nu}$ could not be well justified (think for instance in the {\it hypothetical} production of black holes in particle accelerators).\\

\subsection{Two illustrative $f(R)$ models.}

We will now study the behavior of the function $\phi(T)\equiv f'[\R(T)]/f'[\R(0)]$ for two illustrative models.

\subsubsection{Ultraviolet corrections: $f(R)=R+\frac{R^2}{R_P}$ \label{sec:ultra}}

This model is characterized by a high energy/curvature correction $R^2/R_P$, where the subscript $P$ stands for {\it Planck scale}. In this case, we find that $\R(T)=-\kappa^2T$ is the same as in GR, and the function $\phi(T)$ is given by
\begin{equation}
\phi(T)= 1-2\frac{\kappa^2 T}{R_P}
\end{equation}
We thus see that only at very high matter/energy densities will the function $\phi(T)$ significatively depart from unity. 

\subsubsection{Infrared corrections: $f(R)=R-\frac{\mu^4}{R}$ \label{sec:infra}}

This model was initially proposed in \cite{CDTT04} within the metric formalism and is characterized by a low curvature scale $\mu^2$. In this case, we find that $\R(T)=-(\kappa^2 T+\sqrt{(\kappa^2 T)^2+12\mu^4})/2$ recovers the GR limit for $|\kappa^2 T|\gg \mu^2$ and tends to a constant $\R\sim \mu^2$ for $|\kappa^2 T|\ll \mu^2$. The function $\phi(T)$ is given by
\begin{equation}\label{eq:f'}
\phi(T)=1-\frac{1}{2[1+\sqrt{1+12/\tau^2}]}
\end{equation}
Here $\tau\equiv -T/T_c$, $T_c\equiv\mu^2/\kappa^2\equiv\rho_\mu $, and $\rho_\mu\sim 10^{-26} \ g/cm^3$ represents the characteristic cosmic density scale of the theory, which triggers the cosmic speedup. It is easy to see that at high densities, as compared to $\rho_\mu$, $\phi(T)\to 3/4$, whereas for $\rho\ll \rho_\mu$ we find $\phi(T)\to 1$. Note that we could have chosen the normalization of $\phi(T)$ differently and in such a way that at high densities $\phi(T)\to 1$ whereas at low densities $\phi(T)\to 4/3$. In this latter case, however, the physical metric $g_{\mu\nu}$ would tend to $\frac{3}{4}\eta_{\mu\nu}$ in vacuum. We find our first choice a more natural normalization (though arbitrary anyway), since it makes $g_{\mu\nu}=\eta_{\mu\nu}$ in vacuum.\\

\section{Dirac equation in curved space}

It is well known \cite{Parker80PRL,Parker80,P-P82} that the energy levels of a Hydrogen atom falling freely in an external gravitational field (in GR) will be shifted in a very characteristic way due to the interaction of the electron with the curvature of the space-time. Though external fields in Palatini theories of gravity must also lead to this phenomenon, we will focus here on a different aspect. We will study the effect that the local energy-momentum densities have on the non-relativistic limit of the Dirac equation due to the factor $\phi(T)^{-1}$ appearing in (\ref{eq:phi-eta}).\\
For the sake of clarity, let us briefly consider the different contributions that make up $T$, which generate the metric (\ref{eq:phi-eta}) seen by the system.  The electromagnetic field, which is treated as classical, is traceless and, therefore, does not contribute to $T$. The atomic nucleus can be modeled as point-like or as described by an extremely localized wave-packet contributing with $T_N=-m_N\delta_\epsilon(x)$, where $\delta_\epsilon$ is some representation of the Dirac delta function with spread $\epsilon$ {\it centered} at the origin, and $m_N$ is the nuclear mass. 
The motion of the electron is described by the one-particle Dirac equation, which generalized to curved space-time \cite{Parker80PRL,Parker80,P-P82} can be derived from the following action (the notation will be explained below)
\begin{equation}\label{eq:Dirac-action}
S_m[g_{\mu\nu},\psi]=-\int d^4x\sqrt{-g}\left[i\bar{\psi}\lambda^\mu D_\mu\psi-m\bar{\psi}\psi\right]
\end{equation}
Upon variation of this action with respect to $g_{\mu\nu}$ one finds the energy-momentum tensor associated to the electron, whose trace is given by \cite{B&D} 
\begin{equation}
T_e=-m\bar{\psi}\psi
\end{equation}
In summary, $T=T_N+T_e=-m_N\delta_\epsilon (x)-m\bar{\psi}\psi$.\\
 
\subsection{Derivation of the non-relativistic limit} 
From the action (\ref{eq:Dirac-action}), we can derive the curved space-time version of Dirac's equation 
\begin{equation}\label{eq:Dirac}
(i\lambda^\mu D_\mu -m)\psi=0
\end{equation}
Here $\lambda^\mu=e^\mu_a\gamma^a$ are the curved space Dirac matrices, which are related to the constant Dirac matrices $\{\gamma_a,\gamma_b\}=2\eta_{ab}$ by the vierbein $e^\mu_a$ (recall that $g^{\mu\nu}=\eta^{ab}e^\mu_a e^\nu_b $). The covariant derivative is given by
\begin{equation}
D_\mu=\partial_\mu+ie A_\mu+\frac{1}{2}w_\mu^{ab}\Sigma_{ab}
\end{equation}
with $w_\mu^{ab}$ representing the spin connection, $A_\mu$ is the electromagnetic vector potential, and $\Sigma_{ab}=\frac{1}{4}[\gamma_a,\gamma_b]$. Since, by construction, the matter action is not coupled to the connection $\Gamma^\alpha_{\mu\nu}$, the spin connection $w_\mu^{ab}$ must be defined in terms of the Christoffel symbols $C^\lambda_{\mu\nu}=\frac{g^{\lambda \rho
}}{2}\left(\partial_\mu g_{\rho \nu }+\partial_\nu g_{\rho \mu }-\partial_\rho g_{\mu \nu }\right)$ and the vierbein as $w_\mu^{ab}=e^a_\sigma\nabla_\mu e^{\sigma b}=e^a_\sigma(\partial_\mu e^{\sigma b}+C^\sigma_{\mu\lambda}e^{\lambda b})$. From (\ref{eq:phi-eta}) it is easy to see that $e^a_\mu=\phi^{-1/2}\delta^a_\mu$ and $e_a^\mu=\phi^{1/2}\delta_a^\mu$. After a bit of algebra, (\ref{eq:Dirac}) turns into
\begin{equation}\label{eq:Dirac-1}
\left[i\gamma^a(\partial_a+ieA_a -\partial_a\Omega)-\tilde{m}\right]\psi=0
\end{equation}
where we have defined 
\begin{eqnarray}
\Omega&\equiv& (3/4)\ln \phi(T) \\
\tilde{m}&\equiv& m\phi^{-\frac{1}{2}} \ .
\end{eqnarray}
Even though (\ref{eq:Dirac-1}) is not, in general, completely separable due to the non-linearities introduced by the dependence of $T$ on $\bar{\psi}\psi$, stationary solutions do exist. To find them, it is useful to write the equation in the form $i\partial_t\psi=H\psi$ \footnote{The discussion and construction of the Hilbert space of the solutions of this equation lie beyond the scope of this paper. However, we want to point out certain difficulties related to the fact that the non-linearities induced by the $\psi$-dependence of $T$ are in clear conflict with the superposition principle. 
Note also that the ``Hamiltonian'' $H$ is not hermitian due to the imaginary term $i\nabla\Omega$ (there are also other sources of non-hermiticity which also arise in pure GR and can be neglected when one focuses on the one-particle sector of the theory [see \cite{Parker80} for more details]).}  as follows
\begin{equation}\label{eq:Dirac-2}
i\partial_t\psi=\left[\vec{\alpha}\cdot(\vec{p}-e\vec{A}+i\vec{\nabla}\Omega)+(eA_0+i\partial_t\Omega)+\tilde{m}\beta\right]\psi
\end{equation}
Let us now focus on the positive energy solutions of this equation. It is easy to see that taking $\psi(t,\vec{x})=e^{-iEt}\xi(\vec{x})$ we have $T_e=-m\bar{\xi}\xi$, $\partial_t\Omega=0$, and (\ref{eq:Dirac-2}) turns into
\begin{equation}\label{eq:Dirac-3}
E\xi=\left[\vec{\alpha}\cdot\vec{\pi}+eA_0+\tilde{m}\beta\right]\xi
\end{equation}
where we have used the shorthand notation $\vec{\pi}\equiv (\vec{p}-e\vec{A}+i\vec{\nabla}\Omega)$. Denoting by $\eta$ and $\chi$ the large and small components, respectively, of the Dirac spinor
\begin{equation}\label{eq:large-small}
\xi=\left(\begin{array}{c}\eta \\ \chi\end{array}\right)
\end{equation}
we find the following relations
\begin{eqnarray}\label{eq:chi}
\chi&=&\frac{1}{\tilde{m}+E-eA_0}\vec{\sigma}\cdot\vec{\pi}\eta\\
E\eta&=& \left[\vec{\sigma}\cdot\vec{\pi}\frac{1}{\tilde{m}+E-eA_0}\vec{\sigma}\cdot\vec{\pi}+\tilde{m}+eA_0\right]\eta \label{eq:eta}\\
T_e&=&-m\eta^\dagger\left[I-\vec{\sigma}\cdot\vec{\pi}^\dagger\frac{1}{[\tilde{m}+E-eA_0]^2}\vec{\sigma}\cdot\vec{\pi}\right]\eta \label{eq:T-eta}
\end{eqnarray}
We will now proceed to compute the lowest-order non-relativistic limit. We first decompose the energy $E$ in two parts, $E=m_0+\E$, where $m_0$ is a constant of order $\sim m$ (to be discussed further below) and $\E\ll m_0$ represents the non-relativistic energy. We then expand assuming that the rest mass is much larger than the kinetic and electrostatic energies, $\tilde{m}\sim m_0\gg|\E-eA_0|$, and retain terms only of order $1/m_0$. The above relations reduce to
\begin{eqnarray}\label{eq:chi-1}
\chi&\approx&\frac{1}{\tilde{m}+m_0}\vec{\sigma}\cdot\vec{\pi}\eta\\
\E\eta&\approx& \left[\frac{1}{\tilde{m}+m_0}(\vec{\sigma}\cdot\vec{\pi})^2+(\tilde{m}-m_0)+eA_0\right]\eta \label{eq:eta-1}\\
T_e&\approx&-m\eta^\dagger\left[I-O(|\vec{\pi}|^2/m_0^2)\right]\eta=-m\eta^\dagger\eta \label{eq:T-eta-1}
\end{eqnarray}
The wavefunction of the electron is then identified with $\eta$, which to this order coincides with the positive energy Foldy-Wouthuysen bispinor \cite{Foldy50}. From (\ref{eq:T-eta-1}) we see that the non-linearities contained in $\vec{\pi}$ and $\tilde{m}$ in  (\ref{eq:eta-1}) only depend on $\eta$. Expanding the operator $(\vec{\sigma}\cdot\vec{\pi})^2$ we find
\begin{eqnarray}\label{eq:Pauli}
\E\eta&=& \left\{\frac{1}{\tilde{m}+m_0}[(\vec{p}-e\vec{A})^2-e\vec{\sigma}\cdot\vec{B}]+eA_0\right\}\eta \nonumber\\&+&\left\{\frac{1}{\tilde{m}+m_0}\left[i\vec{\sigma}(\vec{\nabla}\Omega\times\vec{\nabla})-2ie(\vec{A}\cdot\vec{\nabla}\Omega)\right.\right. \\&+&\left.\left.\vec{\nabla}^2\Omega-|\vec{\nabla}\Omega|^2+2(\vec{\nabla}\Omega \cdot\vec{\nabla})\right]+(\tilde{m}-m_0)\right\}\eta  \nonumber
\end{eqnarray}
The first line of this equation is very similar to the well-known non-relativistic Schrodinger-Pauli equation (see (\ref{eq:Sch-Pauli}) below). The only difference being the term $1/(\tilde{m}+m_0)$. The second and third lines, however, represent completely new terms generated by the Palatini gravitational interaction. When the gravity lagrangian is that of GR, $\phi(T)=1$, we recover the Schrodinger-Pauli equation if $m_0$ is identified with ${m}$.\\

\section{Application: the Hydrogen atom}

To gain some insight on the role and properties of the various terms in (\ref{eq:Pauli}), we will proceed as follows. We first solve (\ref{eq:Pauli}) in the case of GR, $f(R)=R, \phi(T)=1$, which is well known. Then we switch to a different gravity lagrangian (assuming that we have the ability to do that) and study how the system reacts to that change. The reason for this is that in a general $f(R)$ the metric is sensitive to the local $T_{\mu\nu}$ via $\phi(T)^{-1}$, and changes in the metric due to the matter distribution could react back on the matter equations. If the new interaction terms in (\ref{eq:Pauli}) lead to small perturbations, then the initial wavefunctions will be, roughly speaking, stable with perhaps small corrections which could be computed using standard approximation methods. If, on the contrary, the energy associated to the gravitationally-induced terms is large, that would mean that the original configuration is not minimizing the modified Hamiltonian and, therefore, large modifications would be necessary to reach a new equilibrium configuration. Depending on the magnitude of the reaction on the system, we could estimate whether the theory is ruled out or not. \\

Let us first consider the $f(R)$ model with ultraviolet corrections introduced in  section \ref{sec:ultra}. In this case, the function $\phi(T)=1-2\frac{\kappa^2 T}{R_P}$ can be expressed as 
\begin{equation}
\phi(T)=1+\frac{2[\rho_N(x)+\rho_e(x)]}{\rho_P}
\end{equation}
where $\rho_N(x)=m_N\delta_\epsilon(x)$, $\rho_e(x)=mP_e(x)$, $P_e(x)=\eta^\dagger(x)\eta(x)$ is the probability density, and $\rho_P\equiv R_P/\kappa^2$ is a very high matter-density scale (Planck scale). Since the scale $\rho_P$ is much larger than any density scale reachable by the electron wavefunction and even by the very peaked nuclear wavefunctions ($\rho_N/\rho_P\sim 10^{-79}$), we see that $\Omega(T)\approx \frac{3}{2}\frac{\rho_N(x)+\rho_e(x)}{\rho_P}$ and $\tilde{m}\approx m(1-\frac{\rho_N(x)+\rho_e(x)}{\rho_P})$ lead to strongly suppressed contributions (in fact, they are much smaller than the corresponding Newtonian corrections $|h_{00}|=GM/c^2R_N\sim 10^{-39}$). Identifying $m_0$ with the electron mass $m$, the leading order corrections to the wave functions and the energy levels could be computed by perturbation methods and would lead to virtually unobservable effects. \\

\begin{figure}[htbp]
\begin{center}
\includegraphics[angle=0,width=3.0in,clip]{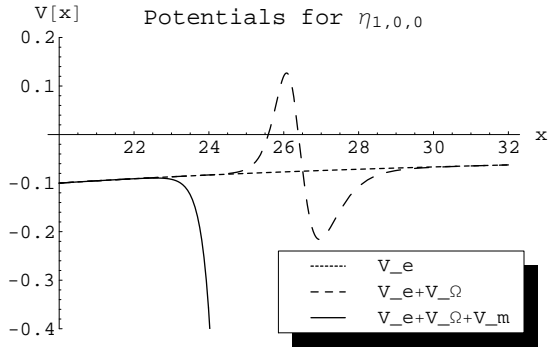}
\caption{Contribution of the different potentials in the ground state. The solid line, which represents the sum of all the potentials, tends to the constant value $-0.134m_0$ (or $-5048$ in the units of the plot).}\label{Fig:ground}
\end{center}
\end{figure}

Let us now focus on the model with infrared corrections introduced in section \ref{sec:infra}. Expressing length units in terms of the Bohr radius ($a_0\sim 0.53\cdot 10^{-10}$m), we find $\tau=\frac{\rho_e(x)}{\rho_\mu}=10^{24} P_e(x)$, where we have intentionally omitted the nuclear contribution (only relevant at the origin) for simplicity. This expression for $\tau$ indicates that the electron reaches the characteristic cosmic density, $\tau\sim 1$, in regions where the probability density is near $P_e(x)\sim 10^{-24}$. In ordinary applications, one would say that the chance to find an electron in such regions is negligible, that that region is empty. In our case, however, that scale defines the transition between the high density ($\tau\gg 1$) and the low density ($\tau\ll 1$) regions.\\ In regions of high density, we find that $\phi$ rapidly tends to a constant, $\phi_\infty=3/4$, which leads to $\tilde{m}=2m/\sqrt{3}$ and $\vec{\nabla}\Omega=0$. If we then identify $m\to \sqrt{3}m_0/2$, equation (\ref{eq:Pauli}) reduces to the usual Schrodinger-Pauli equation
\begin{equation}\label{eq:Sch-Pauli}
\E\eta= \left\{\frac{1}{2m_0}[(\vec{p}-e\vec{A})^2-e\vec{\sigma}\cdot\vec{B}]+eA_0\right\}\eta 
\end{equation}
This fact justifies the introduction of $m_0$ above. Let us now see what happens in regions of low density. In those regions,  $\phi(T)$ tends to unity, $\vec{\nabla}\Omega=0$, and $\tilde{m}\to m$ as $\tau\to 0$. The mass factor dividing the kinetic term is now a bit smaller ($m_0>m$) than in the high density region. But the mass difference $\tilde{m}-m_0$ is not zero. This is a remarkable point, because $\tilde{m}-m_0\approx -0.13m_0$ is negative and of order $\sim m_0$, which represents a large contribution to the Hamiltonian. To better understand the effect of this term, it is useful to consider the ground state, $\eta_{(1,0,0)}=\frac{e^{-r/a_0}}{\sqrt{\pi a_0^3}}\otimes |\frac{1}{2},s\rangle$, where $|\frac{1}{2},s\rangle$ represents a normalized constant bispinor. In this case, the transition from the high density region to the low density region occurs at $r\approx 26a_0$. In Fig.\ref{Fig:ground} we have plotted the most representative potentials in dimensionless form
\begin{eqnarray}
V_e&=& -\frac{2}{x} \label{eq:Ve}\\
V_m&=& \frac{2m_0c^2a_0^2}{\hbar^2}\left[m\phi^{-\frac{1}{2}}-m_0\right] \label{eq:Vm}\\
V_\Omega&=& \left[\frac{2}{x}\partial_x\Omega+\partial^2_x\Omega-|\partial_x\Omega|^2\right] \label{eq:VO}
\end{eqnarray}
where $V_e$ is the electrostatic potential generated by the proton, lengths are measured in units of the Bohr radius, $x=r/a_0$, and energies in units of $\frac{\hbar^2}{2m_0 a_0^2}\approx 13.6$ eV. Note that $V_\Omega=\vec{\nabla}^2\Omega-|\vec{\nabla}\Omega|^2$ only contains the most important contributions associated to $\Omega$.
In this case, $V_\Omega$ represents a small transient perturbation. The mass difference $V_m$, however, introduces a deep potential well in the outermost parts of the atom that must have important consequences for its stability. (Note that this effect is not an artifact of the non-relativistic approximation, since it also occurs in the full relativistic theory (\ref{eq:Dirac-1})-(\ref{eq:Dirac-2}) due to the density dependence of $\tilde{m}$). In the initial configuration of the atom, corresponding to GR, the wavefunction of the ground state is concentrated near the origin, where the attractive electric potential is more powerful ($V_e\to-\infty$). As we switch on the $1/R$ theory, a deep potential well of magnitude $\sim-0.13 m_0$ appears in the outer regions of the atom, where $\rho_e(x)\lesssim \rho_\mu$, which makes the ground state unstable and triggers a flux of probability density (via quantum tunneling) to those regions. The half life of Hydrogen subject to this potential can be estimated using time dependent perturbation theory (see the Appendix) yielding
\begin{equation}\label{eq:halflife}
\tau\equiv\frac{\hbar}{\Gamma}\approx 6\cdot 10^3 s
\end{equation}
We thus see that the initial, stable configuration is destroyed in a lapse of time much shorter than the age of the Universe, which is in clear conflict with experiments \footnote{If to recover (\ref{eq:Sch-Pauli}) in the low density region and to avoid this external potential well we identify $m$ with $m_0$ , we then find a potential barrier of magnitude $\sim +0.13 m$ in the interior of the atom, which makes extremely difficult the capture of the electron by the atomic nucleus and is also in conflict with observations.}. \\

Further evidence supporting the instability of the atom is found in the existence of zeros in the atomic wavefunctions in between regions of high density because, obviously, before (and after) reaching $\rho_e(x)=0$ the characteristic scale $\rho_e(x)\sim \rho_\mu$ is crossed. The  first excited state, $\eta_{(2,0,0)}=\frac{1}{\sqrt{8\pi a_0^3}}(1-\frac{r}{2a_0})e^{-r/2a_0}\otimes |\frac{1}{2},s\rangle$, has a zero at $r=2a_0$. The radial derivatives of $\phi(T)$ at that point are very large and lead to very important perturbations which overwhelmingly dominate over any other contribution (see Fig.2). 
\begin{figure}[htbp]
\begin{center}
\includegraphics[angle=0,width=3.0in,clip]{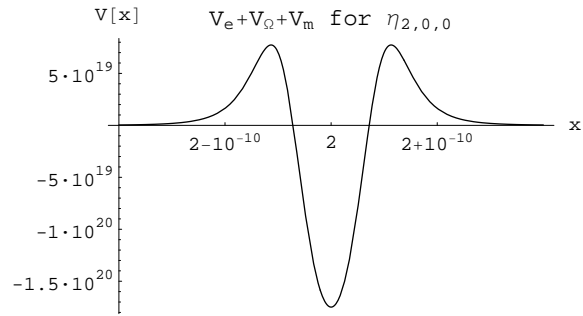}
\caption{The different contributions in this plot are $V_e\sim -1$, $V_m\sim -5\cdot 10^3$, and $V_\Omega\sim \pm 10^{19}$. The y-axis is measured in units of $13.6$eV; the x-axis in units of $a_0$.}\label{fig:2}
\end{center}
\end{figure}
The magnitude of $V_\Omega=\vec{\nabla}^2\Omega-|\vec{\nabla}\Omega|^2$ at $r=2a_0$ oscillates between $10^{20}$ and $-10^{21}$ eV in an interval of only $2\cdot 10^{-10}a_0$. Needless to say that this configuration cannot be stable and that strong changes must take place in the wave function to reduce the energy of the system. Such changes should tend to reduce the magnitude of the density gradients ($\vec{\nabla}\Omega$) to minimize the value of $V_\Omega$, which will likely lead to a rapid transition to the ground state, where $V_\Omega$ is small. One can easily verify that strong gradients $\vec{\nabla}\Omega$ also appear at the zeros of all the $\eta_{n,0,0}$ wavefunctions, which generate  large contributions $V_\Omega$ in those regions. Furthermore, if one considers stationary states with $l\neq 0$, $V_\Omega$ has important contributions not only at the zeros of the radial functions, but also at the zeros of the angular terms. Thus, the pathological behavior described for the spherically symmetric modes gets worse for the $l\neq 0$ states. One thus expects the decay of these states into states with less structure (weaker gradients) such as the ground state,  which will later decay into the continuum.  All this indicates that the existence of bound states, with localized regions of high probability density (where ``high'' means above the scale $\rho_\mu$), are impossible in this theory because of the large gradient contributions $V_\Omega$ and the deep potential well $V_m$. \\

\section{Summary and discussion}

In this work we have deepened into the effects that the matter-energy density dependence of the metric in Palatini $f(R)$ theories has at microscopic scales. In particular, we have studied the effects that switching from GR to a different gravitational interaction, such as the $f(R)=R+R^2/R_P$ or $f(R)=R-\mu^4/R$ models, has on the stationary solutions of the Hydrogen atom. To do so, we started with the Dirac equation in curved space and computed its non-relativistic limit. Then we looked at the contribution of the different new interaction terms appearing in the resulting (effective) Hamiltonian (\ref{eq:Pauli}). We have found that the existence of bound states in theories with infrared corrections is problematic for several reasons. Firstly, due to the dependence of the effective electron mass $\tilde{m}=m\phi^{-\frac{1}{2}}(T)$ on the matter-energy density $T=-m\bar{\psi}\psi$, the effective mass seen in the inner (high density) regions and the outer (low density) regions of the atom is not the same. This generates a potential well that triggers the tunneling of probability density from the inner parts to the outermost parts of the atom, which eventually disintegrates the atom. Secondly, wilder perturbations arise in those points and directions in which the wavefunction has zeros. This is due to the contribution of terms like $\vec{\nabla}^2\Omega$ and $|\vec{\nabla}\Omega|^2$ when the characteristic scale $\rho_\mu$ is crossed \footnote{These contributions represent extreme gradients that cannot be counterbalanced by the electromagnetic interaction to reach new equilibrium configurations. This conflicts with the {\it static configurations} and gradient cancellation claimed in \cite{LMS08} due to their assumption of stability of microscopic systems, which as we have shown here is far from being guaranteed.}. Minimizing the contribution of those terms would require a transition to states with less pronounced gradients such as the ground state, which would latter disintegrate into the continuum via tunneling. \\

Though these instabilities have been discussed within the non-relativistic limit, we do not find any reason to attribute their existence to an artifact of this approximation. In fact, the dependence of the mass on the local energy density was already apparent in (\ref{eq:Dirac-1})-(\ref{eq:Dirac-2}). In addition, derivatives of $\phi(T)$ appear in the term $\partial_\mu\Omega(T)$. Therefore, the relativistic description seems unable to cure the pathologies found in the non-relativistic limit. In addition, one can also check, by direct calculation of  ${R_{\alpha\beta\gamma}}^\lambda{R^{\alpha\beta\gamma}}_\lambda\sim (\partial^2\Omega)^2+\ldots$, that the space-time geometry is strongly fluctuating and far from being flat in those regions where $\rho_e\sim\rho_\mu$. \\
Our results are also likely to hold even in the case in which the spin connection in the matter action is kept independent of the metric. In that case, the connection has a non-vanishing torsion, though the metric $g_{\mu\nu}$ (and hence the vierbein) is still conformally related to the metric $t_{\mu\nu}$ associated to the connection (see \cite{Vol05}), which is the key to get terms of the form $m\phi^{-1/2}$ and $\partial_\mu\Omega$.\\

Though we have only analyzed in detail the infrared-corrected model $f(R)=R-\mu^4/R$, the instabilities associated to the potential well $\tilde{m}-m_0$ and the zeros of the wavefunction must be present in all gravity models sensitive to low curvature/energy-density scales. Since the matter, as we know it, would be unstable in those theories, the cosmological models considered in that context are empty of significance (see \cite{Olmo07} for a list of references). On the contrary, models which introduce deviations from GR at high curvatures, such as $f(R)=R+R^2/R_P$, do not have any relevant effect on the atomic structure if the characteristic scale is sufficiently high. To reach and excite the high energy-density scale one should deal with highly localized wave-packets, which will surely require the consideration of quantum fields. The quantization of the matter fields then opens an exciting window to new phenomena. In fact, when $\psi$ is seen as a quantum field, the function $T$ appearing in (\ref{eq:phi-eta}) and (\ref{eq:Dirac-1})-(\ref{eq:Dirac-2}) must be interpreted as $\langle T \rangle$, i.e., the quantum expectation value of the operator $\hat{T}$ in a given state. The Hamiltonian of the theory then depends on the particular quantum state under consideration through the expectation value $\langle T \rangle$. A direct consequence of this is that the time evolution of the states in the Hilbert space of the theory is nonlinear \cite{Kibble-Weinberg}. This highly non-trivial fact could be used to impose tight constraints on the form of the gravity lagrangian in Palatini theories via quantum experiments. In fact, we believe that in order to guarantee the linear evolution of quantum states, it could be necessary that the gravity lagrangian were exactly that of Hilbert-Einstein.  

\acknowledgements 
The author thanks L.Parker, J. Navarro-Salas, and P.Singh for interesting discussions, R. Parentani for his inquisitive criticisms on earlier versions of this manuscript, and L. Smolin for his hospitality at the Perimeter Institute during the elaboration of this work. This work has been supported by Ministerio de Educación y Ciencia (MEC) and by Perimeter Institute for Theoretical Physics. Research at Perimeter Institute is supported by the Government of Canada through Industry Canada and by the Province of Ontario through the Ministry of Research \& Innovation.

\section*{Appendix}

We briefly sketch here the computation of the half life given in (\ref{eq:halflife}). Our calculation will be approximate and should provide a reasonable estimation of the order of magnitude of $\tau\equiv\hbar/\Gamma$. We will first assume that the kinetic term $-\frac{\hbar^2}{\tilde{m}+m_0}\nabla^2$ can be approximated by $-\frac{\hbar^2}{2m_0}\nabla^2$ everywhere, even though $\tilde{m}+m_0\approx 1.87m_0$ in the low-density regions (recall that $\tilde{m}=m/\sqrt{\phi(T)}$ becomes $m_0$ when $|T|\to\infty$). Secondly,  we will neglect the contribution of $V_\Omega$ and will approximate $V_m(r)=m_0\left[\sqrt{\frac{\phi(\infty)}{\phi(T)}}-1\right]$ by a step function of magnitude $W_S=m_0\left[\sqrt{\frac{\phi(\infty)}{\phi(0)}}-1\right]\approx -0.13 m_0$ in the region $r\ge 26a_0$ and zero elsewhere (see Fig.\ref{Fig:ground}). The total potential (for $l=0$) when the $1/R$ interaction is turned on can thus be seen as 
\begin{equation}
V(r)=\left\{\begin{array}{lr} -\frac{Ze^2}{4\pi\epsilon_0 r} & \makebox{ if } r\leq 26a_0 \\
															-0.13 m_0 & \makebox{ if } r>26a_0\end{array}\right.
\end{equation}															
This way we have reduced our problem to that of an initially stable bound state that becomes unstable and decays into the continuum when the initial potential $U(r)=-\frac{Ze^2}{4\pi\epsilon_0 r}$ is transformed into $V(r)$.  This simplified scenario captures the essential features of our problem.\\
The decay rate can be estimated using time-dependent perturbation theory. A simple and compact expression for the width $\Gamma$ of a quasistationary state (which initially was a true bound state) is given by the following formula (see \cite{Gurvitz} for details)
\begin{equation}\label{eq:decayrate}
\Gamma=\frac{4\hbar^2\alpha^2}{m k}\left|\psi_0(R)\chi_k(R)\right|^2
\end{equation}
In our case, $\alpha=1/a_0$, $k=\sqrt{2m_0(0.13 m_0c^2-|\epsilon|)}/\hbar$, $\epsilon=-13.6 eV$, $\psi_0(R)$ represents the radial part of the partial wave expansion of the ground state evaluated at $R=26a_0$, $\psi_0(R)=\frac{2}{\sqrt{a_0}}\frac{R}{a_0}e^{-R/a_0}$, $\chi_k(r)$ represents the outgoing continuum mode, and $\chi_k(R)=\frac{a_0k}{\sqrt{1+(a_0k)^2}}$. Putting these numbers in (\ref{eq:decayrate}), we find (\ref{eq:halflife}), which implies that the ground state of Hydrogen in the Palatini version of the $1/R$ theory would disintegrate in less than two hours.


\begin{thebibliography}{99}



\bibitem{Tonry-Knop}
J. L. Tonry et al., {\it Astrophys. J.} {\bf 594}, 1 (2003); R. A. Knop
et al., {\it Astrophys. J.} {\bf 598}, 102 (2003).


\bibitem{Olmo07b}
G.J. Olmo, {\it Phys.Rev.} {\bf D} 75,(2007)023511, gr-qc/0612047. 

\bibitem{chameleon}
J. A. R. Cembranos, {\it Phys.Rev.} {\bf D}73 (2006) 064029, gr-qc/0507039;
I. Navarro and K. Van Acoleyen, JCAP 0702 (2007) 022,gr-qc/0611127; 
W.Hu and I.Sawicki, arXiv:0705.1158;
T.Faulkner, M.Tegmark, E.F.Bunn, and Y.Mao, astro-ph/0612569.

\bibitem{Vol03}
D.N. Vollick, {\it Phys. Rev.}{\bf D} 68, 063510 (2003), astro-ph/0306630.

\bibitem{solsys}
X. Meng, P. Wang, {\it Gen.Rel.Grav.} {\bf 36},1947,(2004);
M.L. Ruggiero and L.Orio, {\it JCAP} 0701 (2007) 010, gr-qc/0607093; 
G.Allemandi and M.L. Ruggiero, astro-ph/0610661;
G. Allemandi et al., {\it Gen. Rel. Grav.} {\bf 37},1891 (2005).

\bibitem{Sot06}
T.P. Sotiriou, {\it Gen.Rel.Grav.} {\bf 38} (2006) 1407-1417;
Ph.D. Thesis,  arXiv:0710.4438 [gr-qc].

\bibitem{Fay07}
S.Fay, R. Tavakol and S. Tsujikawa, {\it Phys.Rev.} {\bf D} 75, 063509 (2007);
T.P.Sotiriou, {\it Phys.Rev.} {\bf D}73 (2006) 063515, gr-qc/0509029. 


\bibitem{Olmo07}
G.J. Olmo, {\it Phys. Rev. Lett.} {\bf 98}, 061101 (2007).

\bibitem{Olmo05}
G.J. Olmo, {\it Phys. Rev. Lett.} {\bf 95}, 261102 (2005).

\bibitem{Fla04}
E.E.Flanagan, {\it Phys.Rev.Lett.}{\bf 92}, 071101 (2004).

\bibitem{Sot07}
E.Barausse, T.P.Sotiriou, and J.C.Miller, gr-qc/0703132 (see also arXiv:0712.1141 and arXiv:0801.4852).


\bibitem{LMS08}
B.Li, D.F. Motta, and D.Shaw, arXiv:0801.0603. 


\bibitem{Kai07}
K.Kainulainen et al., arXiv:0704.2729 .

\bibitem{CDTT04}
S.M. Carroll, V. Duvvuri, M. Trodden and M.S. Turner, 
{\it Phys. Rev.} {\bf D} 70, 043528 (2004)


\bibitem{Parker80PRL}
L. Parker, {\it Phys. Rev. Lett.} {\bf 44}, 1559 (1980).

\bibitem{Parker80}
L.Parker, {\it Phys.Rev.} {\bf 22}, 1922 (1980)

\bibitem{P-P82}
L.Parker and L.O. Pimentel, {\it Phys.Rev.} {\bf 25}, 3180 (1982)

\bibitem[Birrell and Davies (1982)]{B&D}
Birrel, N.D. and Davies, P.C.W. (1982). {\it Quantum fields in
curved space}, Cambridge University Press, Cambridge, England.

\bibitem{Foldy50}
L.L.Foldy and S.A. Wouthuysen, {\it Phys.Rev.} {\bf 78}, 29 (1950).


\bibitem{Vol05}
D.N. Vollick, {\it Phys.Rev.} {\bf D} 71, 044020 (2005);
T.P.Sotiriou and S.Liberati, {\it Annals Phys.} {\bf 322} (2007) 935-966.

\bibitem{Kibble-Weinberg}
T.W.B. Kibble, {\it Comm. Math. Phys.} {\bf 64}, 73 (1978);
S. Weinberg, {\it Annals of Physics} {\bf 194}, 336 (1989) 

\bibitem{Gurvitz}
S.A.Gurvitz and G.Kalbermann, {\it Phys.Rev.Lett.} {\bf 59} (1987) 262; S.A.Gurvitz, {\it Phys.Rev.} {\bf A} 38 (1988) 1747.
\end{thebibliography}
\end{document}